\documentclass[11pt]{article}
\usepackage{geometry}                
\usepackage{amsmath}
\usepackage{graphicx}
\usepackage{amssymb}
\numberwithin{equation}{section}
\title{Weyl Symmetry and the Liouville Theory\footnote{Schnitzerfest, Waltham MA, March 2005}
\footnote{Polyakovfest, Princeton NJ, November 2005}\ 
\footnote{Novozhilov Festschrift, ({\it Theoretical Mathematical Physics})}
}

\author{R. Jackiw \\
\it\small Center for Theoretical Physics\\
\it\small Department of Physics\\
\it\small Massachusetts Institute of Technology\\
\it\small Cambridge, MA 02139\\
\\
\small MIT-CTP-3699
}
\date{}                                           

\begin{document}
\maketitle
\thispagestyle{empty}
\begin{abstract}
Flat-space conformal invariance and curved-space Weyl invariance are simply related in dimensions greater than two. In two dimensions the Liouville theory presents an exceptional situation, which we here examine.
\end{abstract}
\newpage

\section{Conformally and Weyl Invariant Scalar Field Dynamics in $d>2$}
Let us begin by recording the d-dimensional Lagrange density for a scalar field $\varphi$ with a scale and conformally invariant self interaction.


\begin{equation}
\mathcal{L}_0 = \frac{1}{2}\ \eta^{\mu \nu}\ \partial_\mu \, \varphi\, \partial_\nu\, \varphi - \lambda\, \varphi^{\frac{2d}{d-2}}
\label{eq:2.1}
\end{equation}
Evidently the expression makes sense only for $d\ne2$, and we take $d>2$. The theory is invariant against 
\begin{equation}
\delta \varphi = f^\alpha\, \partial_\alpha\, \varphi + \frac{d-2}{2d}\ \partial_\alpha\, f^\alpha\, \varphi,
\label{eq:2.2}
\end{equation}
where $f^{\alpha}$ is a (flat-space) conformal Killing vector.
The usual, canonical energy momentum tensor
\begin{equation}
\theta^{\mbox{\scriptsize canonical}}_{\mu \nu} = \partial_\mu\, \varphi\, \partial_\nu\, \varphi - \eta_{\mu \nu}\ \bigg(\frac{1}{2}\ \eta^{\alpha \beta}\, \partial_\alpha \, \varphi\, \partial_\beta\, \varphi - \lambda \, \varphi^{\frac{2d}{d-2}}\bigg)
\label{eq:2.3}
\end{equation}
is conserved and symmetric, as it should be in a Poincar\'{e} invariant theory. But it is not traceless: $\eta^{\mu \nu} \theta^{\mbox{\scriptsize canonical}}_{\mu \nu} \ne 0$.
Nevertheless, because of the conformal invariance (\ref{eq:2.2}), $\theta^{\mbox{\scriptsize canonical}}_{\mu \nu}$ can be improved by the addition of a further conserved and symmetric expression, so that the new tensor is traceless \cite{Callan:1970ze} .
\begin{equation}
\theta_{\mu \nu} = \theta^{\mbox{\scriptsize canonical}}_{\mu \nu} + \frac{d-2}{4(d-1)}\ (\eta_{\mu \nu} \Box - \partial_\mu \partial_\nu)\ \varphi^2, \ \eta^{\mu \nu}\, \theta_{\mu\nu} = 0
\label{eq:2.4}
\end{equation}

A variational derivation of the canonical tensor (\ref{eq:2.3}) becomes possible after the theory (\ref{eq:2.1}) is minimally coupled to a metric tensor $g_{\mu \nu}$, and its action integral is varied with respect to $g^{\mu \nu}$. $\theta^{\mbox{\scriptsize canonical}}_{\mu \nu}$ is regained in the limit $g_{\mu \nu} \to \eta_{\mu\nu}$. A similar derivation of the improved tensor (\ref{eq:2.4}) is also possible, provided (\ref{eq:2.1}), generalized to curved space, is extended by a specific non-minimal coupling \cite{Callan:1970ze}.
\vspace{-1pt}
\begin{eqnarray}
\mathcal{L} = \frac{d-2}{8(d-1)} \ R \varphi^2 + \frac{1}{2}\ g^{\mu \nu}\, \partial_\mu \varphi \partial_\nu \varphi -  \lambda\, \varphi^{\frac{2d}{d-2}} \label{eq:2.5}\hspace{1in}\\[-1ex]
T_{\mu \nu} = \frac{2}{\sqrt{|g|}}\ \frac{\delta}{\delta g^{\mu \nu}}\ \int \, \sqrt{|g|}\, \mathcal{L} \hspace{2in}\nonumber\\[-1ex]
= \partial_\mu \varphi \partial_\nu\, \varphi - g_{\mu \nu} \ \bigg(\frac{1}{2}\ g^{\alpha\beta}\, \partial_{\alpha} \varphi \partial_\beta\, \varphi -  \lambda\, \varphi^{\frac{2d}{d-2}}\bigg)
 + \frac{d-2}{4(d-1)}\ (g_{\mu \nu}\, D^2 - D_\mu D_\nu + G_{\mu \nu} ) \varphi^2 
\label{eq:2.6}
\end{eqnarray}
Here $G_{\mu \nu}$ is  the Einstein tensor, $R$ the Ricci scalar $R = \frac{2}{2-d}\ g^{\mu \nu} \, G_{\mu \nu}$, and $D_\mu $ the covariant derivative. In the limit $g_{\mu \nu} \to \eta_{\mu \nu}$ the  non-minimal term in $\mathcal{L}$ vanishes, but it survives in the $g^{\mu\nu}$ variation.

\begin{equation}
T_{\mu \nu}\ \raisebox{-4pt}{$\overrightarrow{g_{\mu \nu} \to \eta_{\mu \nu}}$} \ \theta_{\mu \nu}
\label{eq:2.7}
\end{equation}
Note that $g^{\mu \nu} \, T_{\mu \nu} = 0$, with the help of the field equation for $\varphi$.
\begin{equation}
D^2 \, \varphi + \lambda \, \frac{2d}{d-1}\ \varphi^{\frac{d+2}{d-2}} -   \frac{d-2}{4 (d-1)} \ R \varphi = 0
\label{eq:2.8}
\end{equation}
This ensures the vanishing of  $\eta^{\mu \nu} \, \theta_{\mu \nu}$.

The precise form of the non-minimal coupling results in the invariance of  the curved space action against Weyl transformations, involving an arbitrary function $\sigma$ \cite{zum70}.
\begin{subequations}\label{eq:2.9}
\begin{eqnarray}
g_{\mu \nu}\ \raisebox{-4pt}{$\overrightarrow{\scriptstyle Weyl}$}\ e^{2\sigma}\, g_{\mu \nu}\label{eq:2.9a}\\
\varphi \ \raisebox{-4pt}{$\overrightarrow{\scriptstyle Weyl}$} \ e^{\frac{2-d}{2}\, \sigma} \, \varphi
\label{eq:2.9b}
\end{eqnarray}
\end{subequations}
The self coupling is separately invariant against (\ref{eq:2.9}). The kinetic term and the non-minimal coupling term are not, but their non-trivial response to the Weyl transformation cancels in their sum. Also it is the Weyl invariance of the action that results in the tracelessness of its $ g^{\mu\nu}$-variation {\it i.e.}\! of $T_{\mu \nu}$, just as its diffeomorphism invariance ensures symmetry and covariant conservation of $T_{\mu \nu}$.

Thus we see that Weyl (and diffeomorphism) invariance in curved space is closely linked to conformal invariance in flat space \cite{zum70}. But can a conformally invariant, flat space theory always be extended to a Weyl and diffeomorphims invariant theory in curved space? Evidently, the answer is ``Yes" for the self-interacting scalar theories in $d>2$, discussed previously \cite{iorio97}. We now examine what happens in $d=2$.

\section{Liouville Theory: Conformally Invariant Scalar Field\\ Dynamics in $d=2$}

A 2-dimensional model with non-trivial dynamics that is conformally invariant is the\\ Liouville theory with Lagrange density
\begin{equation}
\mathcal{L}^{\mbox{\scriptsize Liouville}}_0 = \frac{1}{2}\ \eta^{\mu \nu} \, \partial_\mu \, \psi\, \partial_\nu\, \psi - \frac{m^{2}}{\beta^{2}}\ e^{\beta \psi}.
\label{eq:3.1}
\end{equation}
The conformal symmetry transformations act in an  affine manner, so that the exponential interaction is left invariant.
\begin{equation}
\delta \psi = f^\alpha\, \partial_\alpha \psi + \frac{1}{\beta}\ \partial_\alpha\, f^\alpha
\label{eq:3.2}
\end{equation}
The canonical energy-momentum tensor
\begin{equation}
\theta^{\mbox{\scriptsize canonical}}_{\mu \nu} = \partial_\mu \psi \partial_\nu \, \psi - \eta_{\mu \nu}\ \bigg(\frac{1}{2} \ \eta^{\alpha\beta}\,  \partial_\alpha\, \psi \, \partial_\beta\, \psi - \frac{m^2}{\beta^2}\ e^{\beta \psi} \bigg)
\label{eq:3.3}
\end{equation}
again is not traceless: $\eta_{\mu \nu} \theta^{\mbox{\scriptsize canonical}}_{\mu \nu} \ne 0$, but with an improvement it acquires that property.
\begin{equation}
\theta_{\mu \nu} = \theta^{\mbox{\scriptsize canonical}}_{\mu \nu} + \frac{2}{\beta} \ (\eta_{\mu \nu} \Box - \partial_\mu \partial_\nu)\, \psi , \qquad \eta^{\mu \nu}\, \theta_{\mu\nu} = 0
\label{eq:3.4}
\end{equation}

Again  $\theta^{\mbox{\scriptsize canonical}}_{\mu \nu}$ arises variationally when the Liouville Lagrange density is minimally extended by an arbitrary metric tensor. Similarly the improved tensor (\ref{eq:3.4})  is gotten when a non-minimal interaction is inserted. 
\begin{eqnarray}
\mathcal{L}^{\mbox{\scriptsize Liouville}} = \frac{1}{\beta} \ R \, \psi + \frac{1}{2}\ g^{\mu \nu}\, \partial_\mu \, \psi\, \partial_\nu \psi - \frac{m^2}{\beta^2}\ e^{\beta\psi} \label{eq:3.5} \hspace{2.5in}\\
T_{\mu \nu} = \frac{2}{\sqrt{|g|}}\ \frac{\delta}{\delta g^{\mu \nu}} \, \int \, \sqrt{|g|}\ \mathcal{L}^{\mbox{\scriptsize Liouville}} \hspace{3in} \nonumber\\
= \partial_\mu \, \psi\, \partial_\nu \, \psi - g_{\mu \nu} \bigg(\frac{1}{2}\ g^{\alpha \beta} \, \partial_{\alpha}\, \psi\, \partial_{\beta} \, \psi - \frac{m^2}{\beta^2} \ e^{\beta\psi}  \bigg) 
+ \frac{2}{\beta} \ (g_{\mu \nu} \, D^2 - D_\mu D_\nu )  \psi \label{eq:3.6}\\
 T_{\mu \nu} \ \raisebox{-4pt}{$\overrightarrow{g_{\mu\nu} \to \eta_{\mu \nu}}$} \ \theta_{\mu \nu} \hspace{3.5in}\label{eq:3.7}
\end{eqnarray}
However, the curved-space tensor $T_{\mu \nu}$ is not traceless,
\begin{equation}
g^{\mu \nu}\, T_{\mu \nu} = \frac{2}{\beta^2}\ R \ne 0,
\label{eq:3.8}
\end{equation}
becoming traceless only in the flat-space limit, when $R$ vanishes. Correspondingly, the action associated with (\ref{eq:3.5}) is not invariant against Weyl transformations, which take the following form for the scalar field $\psi$.
\begin{equation}
\psi\ \raisebox{-4pt}{$\overrightarrow{\mbox{\scriptsize Weyl}}$} \ \psi - \frac{2}{\beta}\ \sigma
\label{eq:3.9}
\end{equation}
This formula is needed so that the interaction density $\sqrt{|g|}\ e^{\beta\psi} $ be invariant. However, the kinetic term together with the non-minimal term are not invariant, so that
\begin{eqnarray}
I^{\mbox{\scriptsize Liouville}} &=& \int \, \sqrt{|g|}\ \mathcal{L}^{\mbox{\scriptsize Liouville}}\nonumber\\ 
&&\raisebox{-4pt}{$\overrightarrow{\mbox{\scriptsize Weyl}}$} \ I^{\mbox{\scriptsize Liouville}}  -\frac{2}{\beta^2} \ \int \, \sqrt{|g|}\ (R\, \sigma + g^{\mu \nu}\, \partial_\mu \, \sigma\, \partial_\nu\, \sigma )
\label{eq:3.10}
\end{eqnarray}
Note that the change in the action --- the last term in (\ref{eq:3.10}) --- is $\psi$ independent. So the field equation
\begin{equation}
D^2\, \psi + \frac{m^2}{\beta}\ e^{\beta \psi} - \frac{1}{\beta}\ R =0
\label{eq:3.11}
\end{equation}
enjoys Weyl symmetry, even while the action does not.

\section{Obtaining the $d=2$ Liouville theory from the $d>2$ Weyl invariant theories}
We see that the 2-dimensional situation is markedly different from what is found for $d>2$: for the latter theories there exists a Weyl-invariant precursor, with a traceless energy-momentum tensor in curved space, which leads to a traceless energy-momentum tensor in flat space. For $d=2$ the precursor is not Weyl invariant and the energy-momentum tensor becomes traceless only in the flat-space limit. 

To get a better understanding of the 2-dimensional behavior, we now construct a limiting procedure that takes the Weyl invariant models at $d>2$, (\ref{eq:2.5}), to two dimensions. Thereby we expose the steps at which Weyl invariance is lost.

In order to derive the $d=2$ Liouville theory (\ref{eq:3.5}) from the $d>2$, Weyl invariant models with polynomial interaction (\ref{eq:2.5}), we set 
\begin{equation}
\varphi = \frac{2d}{\beta (d-2)} \ e^{\frac{\beta}{2d} \ (d-2)\ \psi}, 
\label{eq:4.1}
\end{equation}
and take the limit $d\to 2$, from above. We examine each of the three terms in (\ref{eq:2.5}) separately. 

For the self interaction, we have
\begin{equation}
\lambda \varphi^{\frac{2d}{d-2}} = \lambda \bigg(\frac{2d}{\beta(d-2)}\bigg)^{\frac{2d}{d-2}}\ e^{\beta \psi} \ \raisebox{-4pt}{$\overrightarrow{d \downarrow 2}$}\ \frac{m^2}{\beta^2} \ e^{\beta \psi}.
\label{eq:4.2}
\end{equation}
In the last step, to absorb the singular factor  we renormalize the constant $\lambda$ by defining $\frac{m^2}{\beta^2}$. For the kinetic term, the limit in immediate.
\begin{equation}
\frac{1}{2} g^{\mu \nu} \ \partial_\mu\, \varphi\, \partial_\nu\, \varphi \ \raisebox{-6pt}{$\overrightarrow{ d \downarrow 2}$}\ \frac{1}{2}\  g^{\mu \nu}\, \partial_\mu\, \psi \,\partial_\nu \, \psi
\label{eq:4.3}
\end{equation}
But the non-minimal term has no limit, so we expand the exponential.
\begin{eqnarray}
\frac{d-2}{8(d-1)}\ R\, \varphi^2 &=& \frac{d^2}{2\beta^2 (d-1)(d-2)}\ R\, e^{\frac{\beta}{d}\ (d-2)\ \psi} \nonumber\\
&=& \frac{d^2}{2\beta^2 (d-1)(d-2)}\ R + \frac{d}{2\beta (d-1)}\ R \, \psi + \cdot \cdot \cdot
\label{eq:4.4}
\end{eqnarray}
In the $d\!=\!2$ limit, (\ref{eq:4.2}) and (\ref{eq:4.3}) and the last term in (\ref{eq:4.4}) lead to the curved space Liouville Lagrange density (\ref{eq:3.5}). The first term in (\ref{eq:4.4}) gives a indeterminate result in the action.
\begin{equation}
\int \sqrt{|g|} \ \mathcal{L} |_{d>2}\ \raisebox{-4pt}{$\overrightarrow{d\downarrow 2}$}\ \int\, \sqrt{|g|} \ \mathcal{L}^{\mbox{\scriptsize Liouville}} + \frac{2}{\beta^2} \ \frac{\int\sqrt{|g|}\ R}{d-2}
\label{eq:4.5}
\end{equation}
The indeterminancy arises from the fact that in two dimensions $\sqrt{|g|}\ R$ is the Euler density and its integral is just a surface term -- effectively vanishing as far as bulk properties are concerned. So the last term in (\ref{eq:4.5}) gives $0/0$ at $d=2$. Evidently, the Liouville model is regained when $0/0$ is interpreted as $0$, but this leads to a loss of Weyl invariance. To maintain Weyl invariance on the limit $d\downarrow 2$, we must carefully evaluate the $\psi$-independent $\int \sqrt{|g|}\ R/(d-2)$ quantity  -- we need a kind of L'Hospital's rule for dimensional reduction.

It turns out that a precise evaluation of  $\int \sqrt{|g|}\ R/(d-2)$ in the limit $d\downarrow 2$ can be found, by refernce to Weyl's original ideas.

Before describing this, let us remark that the conformal and Weyl transformation rules for $\psi$, (\ref{eq:3.2}) and (\ref{eq:3.9}), are correctly obtained by substituting (\ref{eq:4.1}) into the corresponding rules for $\varphi$, (\ref{eq:2.2})  (\ref{eq:2.9b}), and passing to limit $d\downarrow 2$. The same connection exists between the equations of  motion (\ref{eq:3.11}) (\ref{eq:2.8}). However, the reduction of the $\varphi$ energy-momentum tensor (\ref{eq:2.6}) produces the $\psi$ tensor (\ref{eq:3.6}) plus the term $\frac{4}{\beta^2}\ G_{\mu \nu}/(d-2)$, which is indeterminate at $d=2$, since both the numerator and denominator vanish. Notice that taking the trace of this quantity, before passing to $d\downarrow 2$, leaves $\frac{4}{\beta^2}\ (1-d/2) R/(d-2) = -\frac{2}{\beta^2}\ R$, which cancels the non-vanishing trace of Liouville energy-momentum tensor. This again indentifies the indeterminancy as the source of Weyl non-invariance.

\section{Weyl's Weyl Invariance}
To obtain a definite value for the behavior of $\int \sqrt{|g|} \ R /(d-2)$ in the limit of  $d\downarrow 2$, we examine once again the Weyl transformation properties of the kinetic term for a scalar field theory in $d$ dimensions. (The self-interaction is Weyl invariant, and needs no further discussion.) As already remarked, the kinetic term is not Weyl invariant, and this is compensated by the non-minimal interaction, to produce the Weyl invariant kinetic action.
\begin{equation}
I^{\mbox{\scriptsize kinetic}} = \int \sqrt{|g|}\ \bigg(\frac{1}{2}\ g^{\mu \nu}\, \partial_\mu \, \varphi\, \partial_\nu\, \varphi + \frac{d-2}{8(d-1)}\ R \, \varphi^2 \bigg)
\label{eq:5.1}
\end{equation}

However, Weyl proposed a different mechanism for the construction of a Weyl invariant kinetic term: Rather than using a non-minimal interaction, he introduced a ``gauge potential" $W_\mu$ to absorb the non-variance \cite{iorio97}.
One verifies that
\begin{equation}
I^{\mbox{\scriptsize Weyl}} =  \int \sqrt{|g|}\ \bigg(\frac{1}{2}\ g^{\mu \nu} [\partial_\mu \varphi + (d-2)\, W_\mu \, \varphi] [\partial_\nu\, \varphi + (d-2) \ W_\nu \, \varphi]\bigg)
\label{eq:5.2}
\end{equation}
is invariant against (\ref{eq:2.9}), provided $W_\mu$ transforms as
\begin{equation}
W_\mu\ \raisebox{-4pt}{$\overrightarrow{\mbox{\scriptsize Weyl}}$}\  W_\mu - \frac{1}{2}\ \partial_\mu\, \sigma.
\label{eq:5.3}
\end{equation}

We now demand that $I^{\mbox{\scriptsize kinetic}}$ in (\ref{eq:5.1}) coincideds with $I^{\mbox{\scriptsize Weyl}}$ in (\ref{eq:5.2}). This is achieved when the following holds.
\begin{equation}
\frac{R}{4(d-1)} = D^\mu\, W_\mu + (d-2) \ g^{\mu \nu} \, W_\mu \, W_\nu
\label{eq:5.4}
\end{equation}
this curious Riccati-type equation is familiar in $d=2$, where it states that $\sqrt{|g|}\ R$ is a total derivative; a condition that is generalized by (\ref{eq:5.4}) to arbitrary $d>2$. 

With the help of (\ref{eq:5.4}) we evaluate, before passing to $d\downarrow 2$, the ambiguous contribution to the action --  the last term in (\ref{eq:4.5}). We have from (\ref{eq:5.4})
\begin{equation}
\frac{\int \sqrt{|g|} R}{4(d-1) (d-2)} = \frac{1}{d-2}\ \int \, \partial_\mu\ (\sqrt{|g|} W^\mu) + \int \sqrt{|g|}\, g^{\mu \nu}\, W_\mu W_\nu.
\label{eq:5.5}
\end{equation}
The first term does not contribute, even when $d\ne 2$, because the integrand is a total derivative for all $d$, while the remainder leaves
\begin{equation}
\raisebox{-5pt}{$\overset{\lim}{d\downarrow 2}$}\ \frac{\int \sqrt{|g|} R}{d-2} = 4 \int \sqrt{|g|} \ g^{\mu \nu}\,  w_\mu w_\nu
\label{eq:5.6}
\end{equation}
where $w_\mu \equiv W_\mu |_{d=2}$ satisfies, according to (\ref{eq:4.4}),
\begin{equation}
4 D^\mu w_\mu = R \qquad \mbox{at}\  d=2.
\label{eq:5.7}
\end{equation}
[Note that (\ref{eq:5.3}) and (\ref{eq:5.7}) are consistent with the Weyl transformation formula for $R$ at $d = 2$: $ R\ \raisebox{-4pt}{$\overrightarrow{\scriptstyle Weyl}$}\ e^{-2\sigma}\ (R - 2 D^2\, \sigma)$.]

Thus to achieve Weyl invariance, the action should be supplemented by the metric-dependent, but $\psi$-independent term.
\begin{equation}
\triangle I = \frac{8}{\beta^2} \ \int \, \sqrt{|g|}\ g^{\mu \nu}\, w_\mu w_\nu
\label{eq:5.8}
\end{equation}
According to (\ref{eq:5.3}) and (\ref{eq:5.7}), the Weyl variation of $\triangle I$ is 
\begin{equation}
\triangle I \ \raisebox{-4pt}{$\overrightarrow{\scriptstyle Weyl}$} \ \triangle I + \frac{2}{\beta^2}\, \int \, \sqrt{|g|}\ (R\, \sigma + g^{\mu \nu}\, \partial_\mu \, \sigma \, \partial_\nu \sigma).
\label{eq:5.9}
\end{equation}
This cancels the Weyl non-invariant response of $I^{\mbox{\scriptsize Liouville}}$; see (\ref{eq:3.10}). 

It remains to determine $w_\mu$ by solving (\ref{eq:5.7}). We are of course interested in a local solution, so that the Weyl-invariant Liouville action be local. Such a solution has been found \cite{Deser:1995ne}. It is
\begin{equation}
w^\mu = \frac{\varepsilon^{\mu \nu}}{4\sqrt{|g|}} \ \bigg(\frac{\varepsilon^{\alpha \beta}}{\sqrt{|g|}}\ \partial_\alpha\, g_{\beta \nu} + (\cosh \omega -1)\  \partial_\nu\,  \gamma\bigg).
\label{eq:5.10}
\end{equation}
The second term in the parenthesis is the canonical SL (2, R) 1-form, with 
\begin{equation}
\cosh \omega = \frac{g_{+ -}}{\sqrt{|g|}} \quad \mbox{and}\ e^{\gamma} = \sqrt{\frac{g_{++}}{g_{--}}}.
\label{eq:5.11}
\end{equation}
[$(+, -)$ refer to light-cone components $\frac{1}{\sqrt{2}} (x^0 \pm x^1)$.] 
This portion of $w^\mu$ is Weyl invariant, while the rest verifies the transformation law (\ref{eq:5.3}).
The solution (\ref{eq:5.10}) is not unique. One may add to (\ref{eq:5.10}) any Weyl-invariant term of the form $\frac{\varepsilon^{\mu\nu}}{\sqrt{|g|}} \ \partial_\nu X$, since this will not contribute to (\ref{eq:5.7}). 

Remarkably $w^\mu$ in (\ref{eq:5.10}) is not a contravariant vector, even though $D_\mu w^\mu$ is the scalar $R/4$. Consequently our Weyl invariant Liouville action $I^{\mbox{\scriptsize Liouville}} + \triangle I$ is not diffeomorphism invariant. Its $g^{\mu \nu}$ variation defines a traceless energy-momentuim tensor, which however is not (covariantly) conserved.

We do not know what to make of this. Perhaps the above mentioned ambiguity can be used to remedy the diffeomorphism non-invariance, but we have not been able to do so. It would seem therefore that a local, curved-space Liouville action can be either diffeormorphism invariant or Weyl invariant, but not both.

If this conjecture is true, we are facing an ``anomalous" situation in a classical field theory, which has previously been seen only in a quantized field theory. It is know that in two dimensions, the diffeomorphism invariant Lagrange density $\frac{1}{2} \sqrt{|g|}\ g^{\mu \nu} \partial_\mu \varphi \partial_\nu \varphi$ is also invariant against Weyl transformations that transform the metric tensor, but not the scalar field $\varphi$ [{\it i.e.} Eq. (\ref{eq:2.9}) at $d=2$]. However, the effective quantum action that is obtained by performing the functional integral over $\varphi$, yields a metric expression which is either diffeomorphism invariant or Weyl invariant but not both \cite{polya87}.

If locality is abandoned, one may readily construct a covariant solution for $w_\mu$ in the form $\partial_\mu w$,
\begin{equation}
\omega_\mu = \partial_\mu \omega,
\end{equation}
with $w$ transforming under a Weyl transformation as [compare (\ref{eq:5.3})]
\begin{equation}
w\  \raisebox{-4pt}{$\overrightarrow{\scriptstyle Weyl}$} \ w - \frac{\sigma}{2}.
\label{eq:5.13}
\end{equation}
Evidently
\begin{eqnarray}
D^2 w = R/4, \qquad \qquad \qquad \nonumber\\
w (x) = \frac{1}{4} \ \int d^2 y \ \sqrt{|g (y) |}\ \frac{1}{D^2(x, y)}\  R(y),
\label{eq:5.14}
\end{eqnarray}
where the Green's function is defined by 
\begin{equation}
D^2_x \ \frac{1}{D^2 (x,y)} = \frac{1}{\sqrt{|g|}}\ \delta^2(x-y).
\label{eq:5.15}
\end{equation}
Eq. (\ref{eq:5.13}) is verified by (\ref{eq:5.14}), and the addition to the Liouville action is just the Polyakov action \cite{polya87}.
\begin{equation}
\triangle I = \frac{1}{2\beta^2} \, \int \partial^2 \, x\, d^2 y \sqrt{{|g (x) |}} \ R (x) \ \frac{1}{D^2 (x, y)}\ \sqrt{|g (y) |}\ R (y)
\label{eq:5.16}
\end{equation}
This then provides a diffeomorphism and Weyl invariant action for the  Liouville theory, which however is non-local. Whether locality can be also achieved remains an open question.

\end{document}